\documentclass[preprint,showpacs,preprintnumbers,jap]{revtex4}
\usepackage{amsfonts}
\usepackage{amsmath}
\usepackage{graphicx}
\usepackage{dcolumn}
\usepackage{bm}

\begin{document}

\title{Non-equilibrium dynamics in coupled quantum dots}
\author{H. Cruz \\
{\it Departamento de F\'\i sica, }\\
{\it Universidad de La Laguna, 38204 La Laguna, Tenerife, Spain.}}

\begin{abstract}
The aim of this work is to study the non-equilibrium dynamics of electrons in a coupled quantum well pair. 
To achieve this aim, we consider a non-symmetric distribution of electrons in a double quantum well.
We derive the nonlinear dynamical evolution of the carrier wave functions considering
electron-phonon interactions and a time-dependent Hartree potential in multielectron quantum dots.
We show the possibility of having an electrostatic trap for part of the electrons which
are injected into one of the quantum wells.

{\bf PACS number(s): }73.20.Dx Electron states in low-dimensional
structures, 73.40.Gk Tunneling, 73.50.-h Electronic transport phenomena in
thin films.
\end{abstract}
\maketitle

\section{Introduction}

In the last years, there has been increasing interest in optical and electrical properties of double quantum dots [1-3]. Multielectron quantum dots have attracted much attention motivated by possible applications in quantum computing.
According to Refs. [4] and [5], experimental works on many-electron double quantum dots has demonstrated Pauli blockade and coherent exchange oscillations. In addition to this, differential conductance oscillations were observed in an asymmetric
quantum point contact [6].
In the many-electron quantum regime, Colless {\it et. al.} [7] demonstrated the sensitivity of a sensor, where double quantum dots are commonly operated as charge qubits. 
Such a new device promises to enable readout of qubits in scaled-up arrays. Many-electron qubits typically employ
a Pauli blockade for readout, which requires an exchange rate larger than the tunnel coupling rate.
Multielectron quantum dots are promising candidates for manipulating quantum information [8].

In quantum computing, the ability to control electron trapping with high accuracy plays an essential role.
However, we know that temperature is a common enemy of quantum trapping potentials, i.e., electrons can leave the quantum dot due to phonon-assisted tunneling. 
In this way, we note that phonon-assisted tunneling is a factor that limits the performance of quantum computing [9-11]. Phonon-assisted tunneling is increased as temperature rises. 
The carrier dynamics in a coupled-quantum well pair can be affected by phonon-related 
processes in a thermal environment. In addition to this, 
temperature-assisted transitions between quantum dots can involve charge redistribution when an electron tunnels to a different quantum dot.

One remaining key question is the theoretical analysis of phonon effects on the the efficiency of an electron trap.
A theoretical study of the nonequilibrium dynamics of electrons in a double quantum dot
system can play an important role in quantum computing.
At this point, we note that extending the tunneling analysis to a double quantum-dot considering 
phonon-assited tunneling appears interesting.
Temperature-assisted tunneling in coupled quantum wells is a complex problem which cannot be solved by proposing a unique model to explain experimental observations [9-11]. 
With this is mind, we studied the time-dependent 
evolution of two electron wave packets
in many-electron quantum dots. We know that 
electrical properties can be strongly modified because of electron-phonon interaction in the presence of 
a phonon field. 
If two electron subbands are occupied by
electrons, the electrical properties can also be affected by electron-electron interactions between subbands. Then, we have two carrier groups that interact between each other in a phonon field.

The
method of calculation is based on the discretization of space and time
for both wave functions. We show here that the system dynamics  
can be strongly modified by temperature effects. 
We shall see below that there is a possibility of having an electrostatic trap for part of the electrons localized in one of the quantum dots. 

This paper is organized as follows. Section II describes the model used in this work. In section III the
results are discussed.

\section{Model}

Quantum scattering by longitudinal optical (LO) phonons is summarized in terms of Fermi's Golden rule. In that case, the lifetime of an electron in state
$|i>$ is given by

\begin{equation}
\frac{1}{\tau} = \frac{2 \pi}{\hbar} \sum_{f} |<f|H|i>|^{2} \delta ({\epsilon}_f - {\epsilon}_i)
\end{equation}

where $ {\epsilon}_f$ and ${\epsilon}_i$ are the final and initial energies, respectively. The nature of the lattice vibrations can be described by an angular frequency $\omega$ and a wave vector ${\bold K}$. The
phonon interaction term is obtained by summing over all phonon ${\bold K}$ wave vectors,

\begin{equation}
H = e \sum_{{\bold K}} \left( \frac{\hbar \omega P}{2|{\bold K}|^{2}} \right)^{1/2} \frac{e^{-i {\bold K}{\bold r}}}{V^{1/2}} 
\end{equation}

However, in order to make use of the symmetry of a coupled quantum-well it can be split into components in the $xy$ plane
and along the growth axis. Then, the Eq. (1) becomes [12]

\begin{equation}
\frac{1}{\tau} = \frac{\gamma ''}{2} \Theta \left( k_{i}^{2}- \frac{2 m^{*} \Delta }{\hbar^2} \right) 
\int_{-\infty}^{+\infty} \frac{ \pi |G_{if}(K_{z})|^2 dK_{z}}
{\sqrt{ K_{z}^{4}+2K_{z}^{2}(2k_{i}^2-\frac{2 m^{*} \Delta }{\hbar^2})+(\frac{2 m^{*} \Delta }{\hbar^2})^{2}}}
\end{equation}
where $\Theta$ is a Heaviside unit step function [12], $\Delta = E_{f} - E_{i} \mp \hbar \omega $, where $E_{f}$ and $E_{i}$ represent the energy subband minima of the 
initial (i) and final (f) states, respectively. For parabolic bands, the total energy of a carrier is equal to the energy of the subband minimum plus a component proportional to the momentum squared. The energy $\hbar \omega $ of the LO phonon is only a weak function of the phonon wave vector ${\bold K}$. Hence, the phonon energy $\hbar \omega $ can be approximated with a constant value, taken 36 meV in GaAs. 

\begin{equation}
\gamma '' = 2m^{*} e^{2} \omega P (N_{0}+1/2 \mp 1/2)^/(2 \pi \hbar)^{2}
\end{equation}
and
\begin{equation}
P= \frac{1}{\epsilon_{\infty}}- \frac{1}{\epsilon_{s}}
\end{equation}
with $\epsilon_{\infty}$ and $\epsilon_{s}$ being, respectively, the high and low-frequency
permittivities of the semiconductor.
The Heaviside function ensures that there are only finite lifetimes when 
\begin{equation}
\frac{\hbar^{2} k_{i}^{2}}{2 m^{*}}> E_{f} -E_{i} \mp \hbar \omega.
\end{equation}

The factor $N_{0}+1/2 \mp 1/2$ represents the phonon density within the crystal. The upper sign
of the $\mp$ represents absorption, while the lower sign represents emission of a phonon.
The number of phonons per unit volume is given by the Bose-Einstein factor:

\begin{equation}
N_{0}=\frac{1}{exp(\hbar \omega /k T)-1}
\end{equation}
where $G_{if}(K_{z})$ is know as the form factor and is given by

\begin{equation}
G_{if}(K_{z})= \int \psi_{f}^{*} (z) e^{-i K_{z} z}  \psi_{i} (z) dz.
\end{equation}

The particular form for the wave functions in the z-axis is considered by the form factor $G_{if}(K_{z})$.
At this point, we can notice that it seems reasonable to consider time-dependent wave functions in the 
calculation of the $G_{12}(K_{z})$ form factor. The subscripts $1,2$ refer to electrons in the 
left or right quantum well, respectively (Fig. 1). Then, 
the $G_{12}$ term can be generalized as follows
\begin{equation}
G_{12}(K_{z})= \lim_{T_p \rightarrow \infty} \int_{0}^{T_p} \int \psi_{2}^{*} (z,t) e^{-i K_{z} z}  \psi_{1} (z,t) dz dt
\end{equation}
where $\psi_{1,2} (z,t)$ are the time-dependent wave functions and $T_p$ is a time period. In the coupled quantum-well, the electrons 
are distributed in both quantum states, $\psi_{1}$ and $\psi_{2}$.
In our case, the $\psi_{1,2} (z,t)$ wave functions represent quantum states in two different
quantum dots (Fig. 2).
The $\psi_{1,2} (z,t)$ wave functions in the $z$ axis will be given by the nonlinear Schrodinger equations [13-18]

\begin{equation}
i\hbar \frac{\partial }{\partial t}\psi _{1}(z,t)=\left[ -\frac{\hbar ^{2}}{%
2m^{\ast }}\frac{\partial ^{2}}{\partial z^{2}}+V(z)+V_{H}\left(
\mid \psi _{1}\mid ^{2},\mid \psi _{2}\mid ^{2}\right) 
 \right] \psi
_{1}(z,t),  \label{electron}
\end{equation}

\bigskip\ 
\begin{equation}
i\hbar \frac{\partial }{\partial t}\psi _{2}(z,t)=\left[ -\frac{\hbar ^{2}}{%
2m^{\ast }}\frac{\partial ^{2}}{\partial z^{2}}+V(z)+V_{H}\left(
\mid \psi _{1}\mid ^{2},\mid \psi _{2}\mid ^{2}\right) 
 \right] \psi
_{2}(z,t),  \label{hueco}
\end{equation}%

where $m^{\ast }$ is the electron effective
mass in the conduction band, $%
V(z)$ is the potential given by the double quantum dot, and $V_{H}$ is a Hartree potential due to electron-electron interaction in 
the nanostructure region. 
At this point, we note 
that the Hartree term is a nonlinear potential that
depends on the wave function form.
The Hartree
potential is obtained as follows [13-18]
\begin{equation}
\frac{\partial ^{2}}{\partial z^{2}}V_{H}(z,t)=-\frac{e^{2}}{%
\varepsilon }\left[ n_{1}(t) \left\vert \psi _{1}(z,t)\right\vert ^{2} + n_{2}(t) \left\vert
\psi _{2}(z,t)\right\vert ^{2}\right] ,  \label{poisson}
\end{equation}%
where $n_{s1,s2}$ are the
carrier sheet densities and $\varepsilon $ is the permittivity with respect to the vacuum.

The
carrier sheet densities are given by 
$n_{1,2}=N_{1,2}/A$ are, where $A$ is the sample area.
The $N_1$ and $N_2$ values are the electron number in the left and right quantum dot, respectively.
If an electron tunnels from the left to the right quantum dot, $N_1$ is decreased by $\delta N= 1$ and $N_2$ is increased by $\delta N= 1$. As
a result, the $n_{1,2}$ electron sheet densities are modified by the new electron distribution.

The $N_{1,2}(t)$ values can be obtained with a simple model. 
In our model, the tunneling rate $\Gamma (t)$ for an electron localized in the left dot is given by $ \Gamma (t) = 1/ \tau $, where $\tau = $ $min$ $ \{  \tau_{ph} , \tau_{coh} \} $ is the minimum tunneling time. $\tau_{ph} $ is the phonon-assisted tunneling time and $\tau_{coh} $ is the coherent tunneling time. The transport process is dominated by the shorter tunneling time.

At each time step, the
tunneling rate for an electron in the left quantum well $\Gamma (t)$ can be evaluated. Then, both sheet densities can be easily
obtained.
It is interesting to note that the electron lifetime $ \tau $ depends on the electron sheet densities in the
quantum wells. In that case, every time that an electron tunnels through the barrier, the $\tau (t)$ value is modified.

\section{Results and discussion}

Eqs. (10), (11) and (12) are solved by applying the standard split-step method [13].
The initial probability is set equal to 1.
Eqs. (10), (11) and (12) have been integrated numerically using initial carrier sheet densities equal to $n_1=1\times 10^{11}$ cm$^{-2}$ 
and $n_2=7\times 10^{11}$ cm$^{-2}$.
Fig. 2 shows
the amplitude of both electron wave
functions $\left| \psi _{1,2}\right| ^2$ and conduction band potentials. 
Both electron wave functions are initially created in the center of the left and right quantum wells, respectively, at $t=0$. 
Initial wave packets with zero average momentum have been considered in this work. 
We have considered a GaAs/Ga$_{1-x}$Al$%
_{x}$As double quantum-well system which consists of a left-hand GaAs quantum well (80 {\AA } wide), a right-hand GaAs quantum well (80 {\AA } wide) and a barrier of thickness equal to 20 {\AA }. 
In all numerical examples, $m_{e}^{*}$=0.067$m_{0}$ where $m_{e}$ is the
electron effective-mass.
We solved numerically the Eqs. (10), (11) and (12) using a spatial mesh size of 0.5
\AA ,\ a time mesh size of 1.0 a.u., and a finite box (1350 {\AA }) large
enough so as to neglect border effects.

We obtained the 
probability density of finding carriers, $P_{a,b}$, in a defined semiconductor region [$a$, $b$] by means of the numerical integration over time of Eqs. (10), (11) and (12), 
\begin{equation}
P_{a,b}(t)=\int_a^bdz\ |\psi_{1,2} (z,t)|^2,
\end{equation}
where [$a$,$b$] are the left or right quantum well limits. Fig. 3 shows 
the charge density $|\psi_{1} (z,t)|^2$ in the left-hand quantum well versus time at different electron
densities.
As depicted in Fig. 3,  
the existence of tunneling
oscillations between both quantum wells is clearly shown.
Fig. 3 indicates that  
the amplitude of the coherent tunneling oscillations is approximately equal
to 1 at different charge densities.
The amplitude of the coherent oscillations is strongly reduced at densities higher than 
$n_1=7\times 10^{11} $ cm$^{-2}$ .
This result can be easily explained within the framework of the model.
It is clear that the carrier energy levels of both quantum wells in the conduction band are
exactly aligned at $n_1=0\times 10^{11} $ cm$^{-2}$ and $n_2=0\times 10^{11} $ cm$^{-2}$ (Fig. 1.a). 
In such a case, the initial wave functions, i.e., $\psi_{1} (t=0)$ and $\psi_{2} (t=0)$, will oscillate between both wells with a certain
period owing to coherent tunneling (Fig. 3). 
However, if the potential difference between both quantum dots is different than zero, the resonant condition is not fulfilled and the coherent tunneling period is increased.
If $n_1$ reaches a critical value higher than $n_{1c}=7\times 10^{11} $ cm$^{-2}$,
the tunneling oscillations vanish owing to charge build-up effects.

The different curves plotted in Fig. 3 can be fitted to obtain the tunneling oscillation period. We may assume that the tunneling process is characterized by a coherent tunneling time $\tau_{coh} $.
Fig. 4 shows coherent tunneling time (oscillation period) versus $n_2$ at different $d$ and $n_{1c}$. The critical sheet density value increases as we decrease the barrier thickness.
This effect can be explained as follows. The level splitting (Fig. 1) between both quantum wells depends on the barrier thickness value. If the potential difference between both quantum dots is higher than the level splitting, coherent tunneling is not allowed. 
The level splitting decreases as we increase $d$.
As a result, the critical sheet density takes a lower value if $d$ is increased.
In Fig. 4, we can also notice that coherent tunneling times are decreased as we increase $n_2$. To explain this result, we consider again charge build-up effects. If the right quantum well if filled with carriers, the potential difference between both quantum wells is reduced. Consequently, the tunneling rate between both wells takes a higher value.

Let us now study the phonon-assisted tunneling time $\tau_{ph} $. In Figs. 5 and 6, we plotted $\tau_{coh}$ and $\tau_{ph}$ versus $n_1$ at different $d$ values. 
Examining Figs. 5 and 6, we discover that phonon-assisted tunneling is not allowed below a certain value of the $n_1$ charge density, i.e., $n_1=6\times 10^{11}$ cm$^{-2}$ and $n_1=8\times 10^{11}$ cm$^{-2}$, respectively. 
This effect is given by Eq. (3). The Heaviside function ensures there are only finite lifetimes when energy conservation is considered.
If the barrier thickness takes a higher value (Fig. 6), the potential difference between both quantum wells is reduced due to electron-electron interactions. In such a case, $n_1$ must be increased in order to have phonon-assisted tunneling.

In Figs. 5 and 6, we can also notice that $\tau_{ph}$ increases as we increase $n_1$.
To explain this effect, we take into account that the phonon-assisted tunneling rate is proportional to the energy difference between the initial and final states, see Eq. (3).
The energy difference enlarges with the increase of the $n_1$ charge density. 

As can be seen in Fig. 6, we found a finite gap in the tunneling current between both electron layers at $d=30$ {\AA }. 
In the tunneling gap, the $n_1$ charge density ranges from  
$3 \times 10^{11}$ cm$^{-2}$ to $8 \times 10^{11}$ cm$^{-2}$. 
The tunneling gap depends on $n_1$ and $d$. As a result, we demonstrated the possibility of electrostatic trapping of electrons at certain electron densities.
If the barrier thickness is large enough (Fig. 6), electron tunneling is not allowed at certain $n_1$ values.

To study the practical effects of the tunneling gap, we now simulate the dynamics of electrons which are injected into the device.
Let us now consider the semiclassical dynamics of $N_1$ electrons which are localized in the left quantum well at $t=0$. In principle, the $N_1$ electrons can tunnel through the barrier.
Fig. 7 shows $N_{1} (t)$ versus time at different temperatures. The barrier thickness has been taken to be 20 {\AA}. 

Initially, $N_{1} (t)$ declines gradually owing to phonon-assisted tunneling. Then, the initial decline is followed by a rapid decrease due to coherent tunneling between both quantum dots. This effect is given by the fact that coherent tunneling times are much smaller than phonon-assisted tunneling times.

Examining Fig. 7, we can also notice that electron redistribution between the two wells becomes faster when the temperature is raised. 
This result can be easily explained within the framework of the model: 
the phonon density is increased as we raise temperature as we can see in Eq. (7).

In Fig. 8, we show $N_{1} (t)$ versus time at different $d$ values. We have taken T=77K. 
As can be seen in Fig. 8 ($d=30 $ {\AA} and $d=40 $ {\AA}), part of the electrons remain trapped in the left dot as time progresses.
Such an effect is given by the existence of a finite gap in the tunneling current (Fig. 6). 
If the barrier thickness is large enough, part of the electrons remain localized in the initial state. 
An electrostatic trap is then found to be possible for part of the electrons at $d=30 $ {\AA} and $d=40 $ {\AA}.

Finally, we think that this effect can be observed experimentally. In Fig. 9 is shown a schematic illustration of the proposed experiment.
It consist of a double quantum dot with an adjacent quantum point contact (QPC), which serves as a detector. In these systems, quantum point contacts are used to detect the charge configuration in a coupled quantum-well pair.
One of the quantum dots is coupled to an electron reservoir at $t=0$. 
The conductance of the QPC is sensitive to the occupancy of the right quantum dot.
The quantum point contact destroys the quantum coherence in a double quantum dot and localizes each electron in one of both
dots according to Von Neumann's postulate, consistently with the measurement outcome. 
Electrons are projected onto a well define quantum dot after the observation takes place, if the two quantum wells
are highly isolated.
Electrons are injected into the left quantum dot from an electron reservoir at $t=0$. The current through the QPC directly measures the position of the electron in the right quantum well.
In such a case, the tunneling amplitude in the detector is $\omega + N_2 \delta \omega $ when the right dot is occupied. $ \omega $ is the barrier height in the QPC. 

In summary, in this work we numerically integrated over space and time two effective-mass Schr\" odinger equations in a bilayer electron system considering many-body interactions.
Phonon-assisted tunneling was also considered in our calculations through a time-dependent model.
It is found that the nonlinear electron dynamics in the bilayer are determined by two competing processes: coherent tunneling and phonon-assisted tunneling. If the barrier thickness exceeds a critical value, 
we obtained a finite gap in the tunneling current between both electron layers.
In this way, we showed the possibility of an electrostatic trap for part of the electrons at certain barrier widths.

\newpage

\section{Figures}

\begin{itemize}
\item {\bf Fig. 1} Energy band diagram of the coupled quantum well. (a) Band diagram in absence of electrons. (b) We took an initial sheet density $n_1=7\times 10^{11}$ cm$^{-2}$ and $n_2=0\times 10^{11}$ cm$^{-2}$.

\item {\bf Fig. 2} Conduction band potential and carrier wave
functions at $t=0.4$ ps. We took an initial 2D carrier sheet density
equal to $n_{1}=1\times 10^{11}$ cm$^{-2}$ and $n_{2}=7\times 10^{11}$ cm$^{-2}$.
We considered a GaAs/Ga$_{1-x}$Al$%
_{x}$As double quantum-well system which consists of two 80
{\AA } wide GaAs quantum wells separated by a 20 \AA barrier. 

\item {\bf Fig. 3} Probability density $|\psi_{1} (z,t)|^2$ in the left quantum well ($P_{ab}$) versus time. We took an initial 2D electron sheet density equal to $n_{1}=1\times 10^{11}$ cm$^{-2}$ and 
$n_{2}=7\times 10^{11}$ cm$^{-2}$. Thin line: $n_{1}=1\times 10^{11}$ cm$^{-2}$. Thick line: $n_{2}=7\times 10^{11}$ cm$^{-2}$. We 
considered the double quantum-well system described in the caption of Fig. 2.

\item {\bf Fig. 4} Coherent tunneling time
versus $n_2$ at different $d$ values.

\item {\bf Fig. 5} Tunneling time versus $n_1$. We have taken a barrier with thickness of 20 \AA.

\item {\bf Fig. 6} Tunneling time versus $n_1$. We have taken a barrier with thickness of 30 \AA.

\item {\bf Fig. 7} $N_1$ versus time at different temperatures. We have taken a barrier with thickness equal to 20 \AA.

\item {\bf Fig. 8} $N_1$ versus time at different $d$ values. The temperature is 77K.

\item {\bf Fig. 9} A schematic illustration of the proposed experiment.

\end{itemize}

\end{document}